\documentclass{aa}
\usepackage{capt-of}

\usepackage{multirow}
\usepackage[normalem]{ulem}

\usepackage{graphicx}
\usepackage{txfonts}

\usepackage{capt-of}

\usepackage[colorlinks=true,linkcolor=blue,citecolor=blue,urlcolor=blue]{hyperref}

\newcommand{\lamand}{$\lambda$\,Andromedae}
\newcommand{\lama}{$\lambda$\,And}
\newcommand{\cahk}{\ion{Ca}{ii}\,H\&K}
\newcommand{\cairt}{\ion{Ca}{ii}\,IRT}
\newcommand{\cairtt}{\ion{Ca}{ii}\,IRT~8542~$\rm\AA$}
\newcommand{\Halpha}{H$\alpha$}

\newcommand{\ms}{m\,s$^{-1}$}

\begin{document}

\title{Forward modeling solar spectra onto Doppler images of \lama}

 \author{\"O. Adebali\inst{1, 2} 
 \and A.G.M.\ Pietrow\inst{1} 
 }

 \institute{\inst{1}Leibniz-Institut für Astrophysik Potsdam (AIP), An der Sternwarte 16, 14482 Potsdam, Germany\\
 \inst{2}Institut f\"ur Physik und Astronomie, Universit\"at Potsdam, D-14476 Potsdam, Germany\\
       \email{oadebali@aip.de, apietrow@aip.de}\\
      }

\date{Receieved: 12 February 2026 / Accepted: 27 April 2026}

  \abstract
  % context heading (optional)
  % {} leave it empty if necessary  
  %
   {
   %Large spots influence the spectral and photometric variability of stars. 
   Due to their high chromospheric activity and photometric variability, RS\,CVn type binaries are ideal laboratories for studying stellar surface structures and the corresponding stellar activity relations. However, the atmospheric nature of their primary evolved components (luminosity classes III–IV) are more complex than those of the main-sequence stars. Additionally, the detailed models are still lacking for the sub(giant) systems. Therefore, comparative techniques represent the most effective approach for probing the connection between chromospheric emission and surface structures.
   }
  % aims heading (mandatory)
  %
   {By using the Doppler images of \lama, we aim to investigate whether surface temperature information can be reversed to create its activity parameters, by feeding a toy model with solar spectra, based on the surface images. At the same time, we examine whether spot contributions alone are sufficient to explain the observed activity modulation of the RS\,CVn star \lama\,while quantifying the differences with the actual observations of this star that are obtained simultaneously with the Doppler images we use.
    }
  % methods heading (mandatory)
  %
   {Due to a lack of publicly available starspot models for its stellar type, we adopt observed solar spectra as the only available approximation of \lama's spots. These spectra are injected into a sequence of full-disk temperature map derived from Doppler imaging that represent a full stellar rotation. These disks are then forward modeled into disk-integrated spectra with the Numerical Empirical Sun-as-a-star Integrator (NESSI). This experiment is performed on three photospheric lines (Fe {\sc i} 6173~\AA, Fe {\sc i} 6301~\AA, K {\sc i} 7699~\AA) and four chromospheric lines (\Halpha, \cahk, and Ca {\sc ii} 8542~\AA). Finally, these spectra are used to calculate the radial velocities and chromospheric emissions diagnostics, which in turn are compared to the original photospheric and chromospheric characteristics of the star.
  }
  % results heading (mandatory)
  %
   {Despite the very different stellar structures and atmospheric stratification between \lama\,and the Sun, we show that the chromospheric emissions produced by our toy model largely follow the same trend as the original observations of \lama. This indicates that the modulation of the chromospheric activity is dominated by magnetic activity associated to the active regions with dark spots. In addition, the differences in the emission amplitudes quantify the different chromospheric heating mechanisms for these two very different types of stars.  
    }
  % conclusions heading (optional), leave it empty if necessary 
   {Using this approach, we show that even with simplified assumptions the spectral behavior of \lama\,can be qualitatively reproduced. Toy models such as the one presented in this work procure an additional dimension, providing a relation between the surface structures and chromospheric emissions. It also helps to develop a further understanding for the heating mechanisms of these active giants through comparative techniques, where in this case the spot activity seemingly modulates the chromospheric signal and can explain the bulk of its variations over a rotation.}


 \maketitle
%
%________________________________________________________________

\begin{table*}
\captionof{table}{Summary of SST observations used in this study.}
\label{tab1}
\centering
\begin{tabular}{lccccclcc}
\hline\hline
Archive ID & NOAA & Date (UTC) & X (") & Y (") & $\mu$ & $T_{\rm U}$ (K) & $T_{\rm P}$ (K) & Spectral lines \\
\hline

\href{https://dubshen.astro.su.se/sst_archive/observations/456}{456} &
13395 &
2023-08-06 08:51--08:58 &
-655 &
\phantom{-}169 &
0.71 &
4581 &
6168 &
\begin{tabular}[c]{@{}l l@{}}
K\,I     & 7699~\AA\\
Fe\,I    & 6173~\AA\\
Ca\,II\phantom{\,K} & 8542~\AA
\end{tabular}
\\ \hline

\href{https://dubshen.astro.su.se/sst_archive/observations/416}{416} &
13433 &
2023-09-15 08:38--09:08 &
-454 &
\phantom{-}355 &
0.80 &
4517 &
6078 &
\begin{tabular}[c]{@{}l l@{}}
Fe\,I    & 6301~\AA\\
Ca\,II\,H & 3968~\AA\\
Ca\,II\,K & 3933~\AA
\end{tabular}
\\ \hline

\href{https://dubshen.astro.su.se/sst_archive/observations/423}{423} &
13468 &
2023-10-20 09:05--10:51 &
-380 &
-247 &
0.88 &
4549 &
6197 &
\begin{tabular}[c]{@{}l l@{}}
H$\alpha$\phantom{K I\,} & 6563~\AA
\end{tabular}
\\ \hline

\hline
\end{tabular}
\end{table*}

\section{Introduction} 
Spots are among the most easily identifiable examples of magnetic field inhomogeneities in the stellar atmospheres. They are characterized by a dark umbra surrounded by a brighter, but still dark, penumbra \citep{Solanki2003}.
The spectral profiles of spots are shaped by markedly different local thermodynamic and magnetic conditions \citep[e.g., ][]{ Avrett2015, Kuckein2021}. Nevertheless, when modeling spots on stars other than the Sun, these structures are usually reduced to uniform surface elements that share the same atmospheric stratification as the quiet photosphere, aside from a decrease in effective temperature. \citep[e.g.][]{Chakraborty2024, Petit2024, Cristo2025}. Conversely, state-of-the-art magnetohydrodynamic simulations of starspots are currently being developed \citep[e.g.][]{Smitha2025}, but not yet widely implemented.

Studying these features provides insight into the connection between the activity of different atmospheric layers, such as; the photosphere and the chromosphere \citep{Strassmeier2009}. The Sun displays relatively small spots, typically up to about 100 MSH\footnote{Millionths of a solar hemisphere.}, whereas spots several orders of magnitude larger and filling factors of tens of percent have been reported on other stars. However, it is unclear whether these are singular monolithic spots or rather clumps of smaller unresolved spots \citep{Solanki2004}. 

The primary components of RS CVn type binary systems in particular tend to have very large starspots and filling factors, as well as excessive chromospheric activity. These characteristics make them appropriate targets for spot detection and observation. These systems have been extensively studied over the past fifty years, beginning with \citet{Hall1972} and later, through the development of the Doppler imaging technique \citep{Vogt1983}, which provided a more detailed approach to study these giant activity laboratories. In the following years, subsequent studies have provided more information on different systems, enabling us to understand various aspects of stellar activity such as; stellar differential rotation, activity cycles, magnetic fields and stellar winds \citep[see e.g.,][]{Strassmeier2009, JAG2016}. 

However, these higher activity levels present their own limitations. For example, it becomes more difficult to obtain more precise and accurate results for stellar parameters such as mass and radius. Radial velocity (RV) is one such parameter that is highly affected by large spots, which can contaminate and even mimic the RV signal originating from orbiting exoplanets \citep[e.g., ][]{Moulds2013, Bortle2021, Simpson2022}. Additionally, other activity features, such as flares \citep{Reiners2009, Pietrow2024_harps} and faculae \citep{Cristo2025} have also been shown to create perturbations in RV signals.

\lamand\,(\lama) is a famous RS CVn system displaying massive spots, which has been studied for over a century \citep[][]{Donati1995, Parks2021, Fionnagain2021}. In a recent work by \citet{Adebali2025}, enhanced chromospheric activity was investigated, where they showed varying activity modulation from 10\% to 50\% within a time span of $\simeq$ 10 rotations, for different activity indicators such as; \cahk, \Halpha\,and the Ca {\sc ii} infrared triplet (IRT). Moreover, the RV analysis shows a long-term modulation with an amplitude of 300 \ms. 

Although many of the aforementioned techniques provide detailed results for inferred quantities of \lama, more detailed and comparative methods are lacking to understand the activity imprints of its surface structures. For this reason, it is important not only to employ different techniques, but also to test them in robust ways. In this work, we plan to do exactly this by creating a toy model of \lama\,based on the Doppler imaging map from \citet{Adebali2025}, which in turn can be forward modeled into a spectral time series for different lines to be used for further activity analysis. We note that the surface images of \lama\,were obtained from a data set with very high spectral resolution (R~$\sim$~250\,000) and signal-to-noise ratio (reaching up to 1000). The observations cover a total of 40 days for the primary star of the system ($\rm P_{rot} =$ 54 days) with well-distributed sampling. We therefore assume that despite the small rotational broadening, the reconstructed surface structures represent the magnetic activity of the star reasonably well \citep[see details in][]{Adebali2025}.

To achieve this, we used sunspot spectra obtained from different regions of the sunspots and their surrounding areas. These spectra were then used as input for the Doppler images. In practice, we mapped the temperature distribution derived for \lama\,onto the solar surface and searched for resulting differences in the activity indicators. A more detailed description of this procedure is provided in Section \ref{sec:Nessi}.

Traditional approaches have been used to model activity indicators such as \cahk\,and \cairt\,for different stars by employing stellar template spectra \citep[][]{Mittag2013, Martin2017}. However, these methods may fail for RS\,CVn systems for two main reasons. First, the determination of stellar parameters for RS\,CVn stars, such as visual magnitudes, is often uncertain. Because the primary components of these systems exhibit large starspots, their observed stellar parameters can be significantly affected by stellar activity \citep[see][]{Adebali2026}. Second, these approaches often rely on simplified approximations when modeling chromospheric emission \citep[e.g. ][]{Han2025}. In this work, we address this limitation by using direct observations of spots from the Sun to represent chromospheric contributions.

In the following sections, we first introduce our stellar-disk modeling, and then explain our activity analysis both in photospheric, via RV regime, and chromospheric layers. In the conclusions, we investigate the differences between the activity signatures created by our toy model and the actual observations of \lama.

\section{Stellar disk modeling with \texttt{NESSI}} \label{sec:Nessi}

This toy model is generated using the Numerical Empirical Sun-as-a-star Integrator \citep[NESSI, ][]{Pietrow2024, dewilde2025}, which employs a radial polar sampling scheme that traces a one-dimensional spiral across the stellar disk. Each point along this spiral corresponds to a surface element with a specific area to which an appropriate spectrum can be assigned. NESSI functions in a similar manner to the Spot Oscillation And Planet \citep[SOAP;][]{Cristo2025} code, but it is optimized for detailed surface features and empirical data, making it ideal for this use-case.

NESSI works by selecting a 'disk center' spectrum that spans a given spectral range. This spectrum is then modulated with a wavelength-dependent limb-darkening curve such as those given by \citet{Pietrow2023}, \citet{Canocchi2024},  and \citet{Ellwarth2023}. Then, a rotational profile, as described in \citet{Emily2025} is applied to the disk at a given heliographic latitude\footnote{This is equal to the inclination of rotation value but with the zero defined at the solar disk center. Meaning that $B_0 = 90^\circ - i$, where $B_0$ is the heliographic latitude, and $i$ the inclination of rotation} (in this case, 20$^\circ$), and a rigid rotation of 7.0 km/s is introduced \citep{Adebali2025}, as the differential rotation rate is assumed to be negligible for this star, and the Doppler method is not sensitive to differential rotation over just one rotation. Thus, the main reason why we do not consider differential rotation is that our observations do not cover multiple rotations, contrary to the case of e.g., KU Peg \citep{Kovari2016}.

For this work, we slightly modified the method, as we use the Doppler imaging map from \citet{Adebali2025} to create intensity-based masks that separate the map into three thresholds, namely quiet-Sun, penumbra, and umbra. We apply these maps to the NESSI grid and fill each of the three masks with spectra of the corresponding regions, as explained below. The grid is then processed normally with limb-darkening and rotation. Here, we assume that spots and the quiet Sun have the same limb-darkening curves, despite the evidence suggesting otherwise \citep[e.g., ][]{Rodberg1966,Cretignier2024}. However, currently, no empirical or synthetic limb-darkening atlases exist for sunspots besides a few individual lines. Based on preliminary results from Pietrow et al. (in prep.), we estimate that the difference in limb darkening remains below 5\% for the majority of the time a sunspot spends crossing the disk. Only within the last $\sim$2\% of the solar radius near the limb does this discrepancy increase to around 15\%. For this reason, we consider the effect to be negligible compared to the uncertainties introduced by other assumptions in this work.

%---------------------------------- Fig. 1
\begin{figure}[t]
\includegraphics[width=0.5\textwidth]{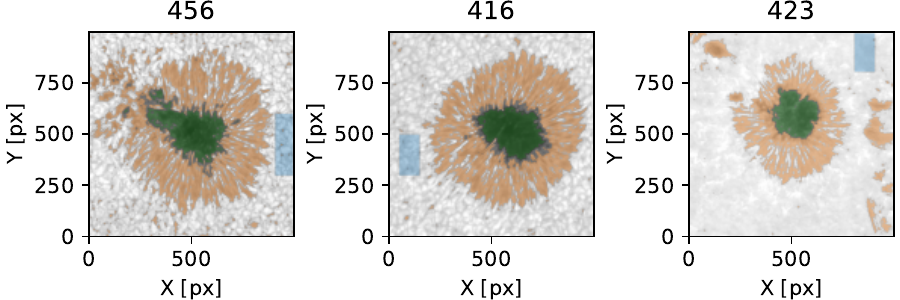}
\captionof{figure}{High resolution sunspot images corresponding to the datasets shown in Table \ref{tab1} with highlighted quiet Sun (blue), penumbra (orange), and umbra (green) regions obtained with intensity thresholds. Pixels within highlighted regions are averaged into their respective spectral profile. The images are of the continuum left of Fe I 6173 \AA, Fe I 6302 \AA, and the H$\alpha$ pseudo-continuum. All figures are observed at a resolution of 0.058 "/pixel.}
\label{fig:contour}
\end{figure}

At present, neither the Sun nor other stellar types have a publicly available database of physically consistent synthetic spot spectra. Empirical spot spectra are also rare and fragmented, with the only available resource being a low signal-to-noise umbral atlas spanning wavelengths from 6642~\AA\ to 11 230~\AA\ \citep{Wallace1999}. However, a self-consistent sunspot atlas is expected to become available in the near future with the commissioning of the Paranal solar ESPRESSO Telescope \citep[PoET;][]{Santos2025}. Until then, we use the next best thing in the form of empirical spot spectra observed with the Swedish 1-m Solar Telescope \citep[SST;][]{Scharmer2003}. These observations offer a high spatial ($\sim$ 0.058 arcseconds/pixel) and spectral (R $\sim$ 150,000) resolution over a relatively small spectral range of about 1 \AA. The data used come from the publicly available SST archive\footnote{\url{https://dubshen.astro.su.se/sst_archive/}} and are observed with both the CRisp Imaging SpectroPolarimeter \citep[CRISP, ][]{Scharmer08} and the CHROMospheric Imaging Spectrometer \citep[CHROMIS, ][]{Scharmer17}. Unfortunately, there is no single sunspot observation that covers all the spectral lines of interest. For this reason, multiple observations were combined (see Table~\ref{tab1} and Fig.~\ref{fig:contour}). As a result, our spectra do not sample the same sunspot regions and were recorded at different locations on the solar disk. While this introduces obvious limitations in terms of self-consistency, such as variations in spectral width and depth with heliocentric angle $\mu$ \citep[e.g.][]{Pietrow2023, Pietrow2025}, we argue that the resulting average umbral and penumbral spectra are still an adequate approximation for the present study. This is because Doppler shifts can be removed, intensities (of the order of 10\% due to limb darkening) can be rescaled, and the remaining profile differences are expected to behave smoothly with $\mu$ between the different types of spectra.
The intensity calibration was performed using the \texttt{get\_calibration} function from the \texttt{ISPy} package \citep{DiazBaso2021}. This routine fits the continuum, where available, or otherwise the far wings of the spectral line using the spectral atlas of \citet{Neckel1984} applied to the empty-disk observations, and subsequently transfers this calibration to all remaining spectra. Before fitting the atlas, spectra are convolved with the SST instrumental profile and include a rotational broadening term following Eq.~18.14 of \citet{Gray2022}. The line core is excluded from the fit due to contamination by activity-related signals, which are present in this and most other atlases \citep{Hanassi2025,Pietrow2026}. Additionally, by comparing the blackbody, or brightness, temperatures of the three spots in the 6173 \AA\,continuum, using data from the Helioseismic and Magnetic Imager \citep[HMI;][]{Scherrer2012} onboard the Solar Dynamics Observatory \citep[SDO;][]{Pesnell}, we find that their temperatures are very similar. The median umbral brightness temperatures of each spot cluster within about 30\,K of 4550\,K, while the penumbral temperatures fall within roughly 100\,K of 6150\,K. The Quiet Sun surface temperature at this height in the photosphere is 6700\,K \citep[e.g.,][model C]{Fontenla1993}.

\subsection{Spectral averaging and NESSI maps}

For each sunspot observation, we manually select an intensity threshold that can be used to split the image into umbra, penumbra, and quiet Sun pixels (see Fig~\ref{fig:contour}), which are then averaged and shifted to remove any Doppler shifts (see Fig~\ref{fig:avgprof}). 

\begin{figure}[t]
    \centering
    \includegraphics[width=\columnwidth]{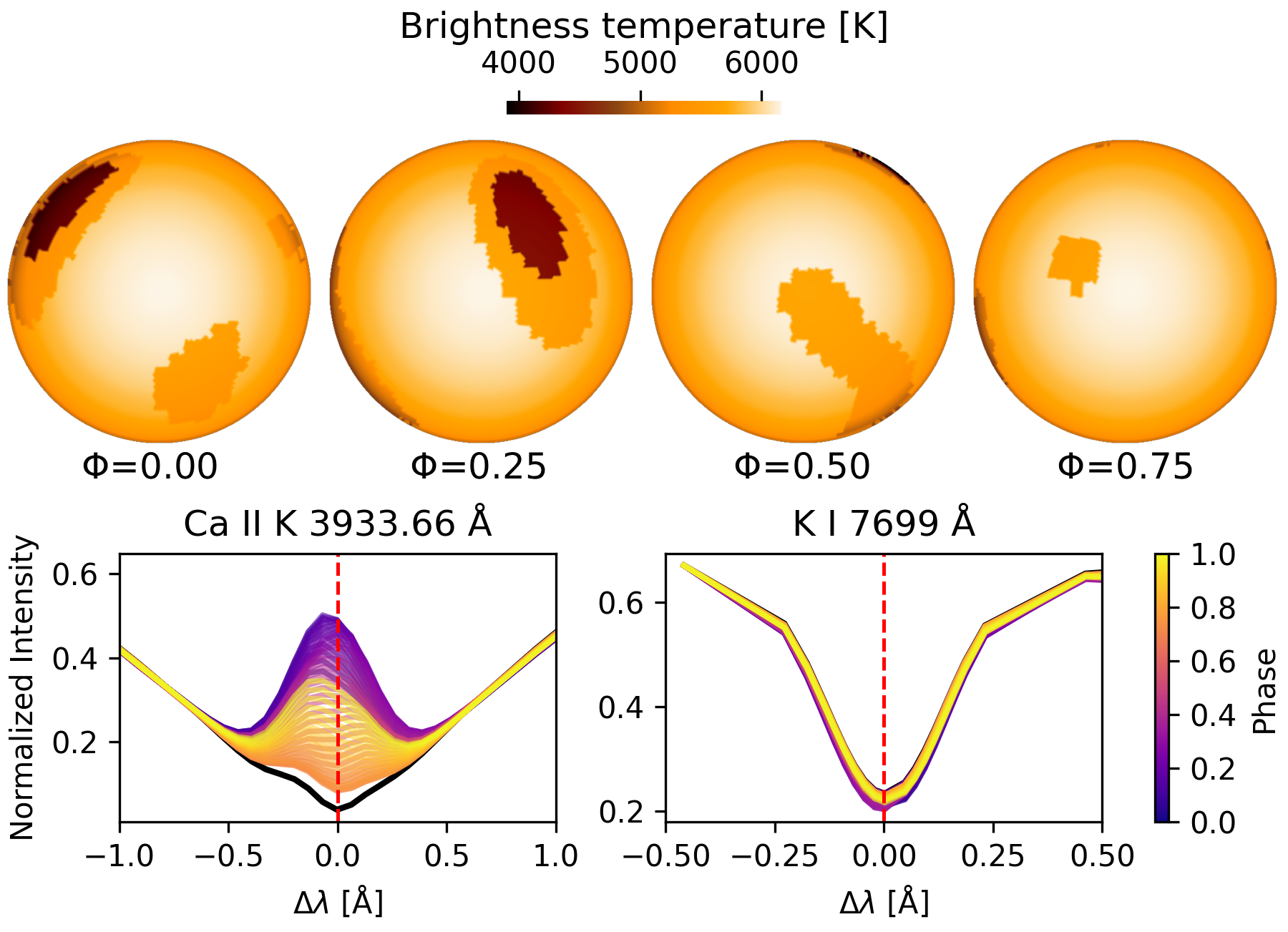}
    \caption{Simulated stellar disks and their profiles. Top: Four resolved NESSI disk brightness temperature maps based on the provided Doppler imaging map with distinct umbra, penumbra, and quiet Sun regions, as well as applied limbdarkening. From left to right, the phases are 0, 0.25, 0.5, and 0.75. Bottom: Resulting spectra for Ca {\sc ii} K and K {\sc i} 7699~\AA\ for all phases (colored) and the disk with no spots (black). The red dashed lines denote the rest wavelength of the line.}
    \label{fig:nessires}
\end{figure}

These profiles are then injected into the NESSI disk, as explained above, and the process is repeated for the 60 masks created from the Doppler imaging map for each line. These grids were then numerically integrated to create synthetic stellar spectra of the disk in question (see Fig.~\ref{fig:nessires}). An 'empty' (non-spotted) disk is created for each line to use as the 'null-spectrum' in the Doppler and RV analysis. 

\section{Activity analysis} \label{sec:Activity}

%---------------------------------- Fig. 2

To compare activity levels, we examined changes in line-core intensity and radial velocity calculations over one rotation of our toy model.   
First, we examined how surface structures affect the emissions at different chromospheric layers. Then, to make a broader comparison with the actual observations, we calculated the activity-induced RV changes; that is, the activity imprints at the photospheric layer.

%---------------------------------- Fig. 3
\begin{figure*}[t]
    \includegraphics[width= \textwidth]{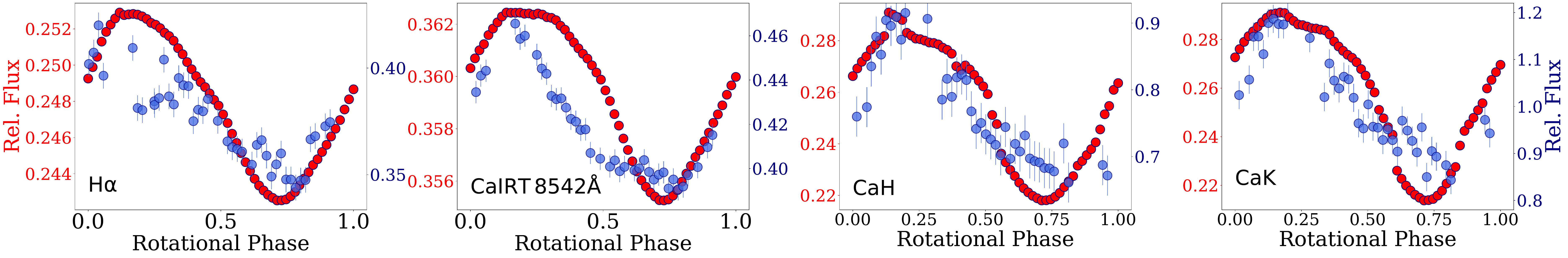}
    \caption{Chromospheric emission variability. Emissions from \Halpha, Ca {\sc ii} 8542~\AA, and \cahk\,are plotted versus rotational phase of the star as modeled (red) and observed (blue) values. The scale differences between the modeled and observed values are indicated with red and blue y-axis colors respectively. The emissions are plotted in fluxes relative to the continuum.}
    \label{Chromo_figs}
\end{figure*}

%---------------------------------- Fig. 4
\begin{figure*}[t]
    \includegraphics[width= \textwidth]{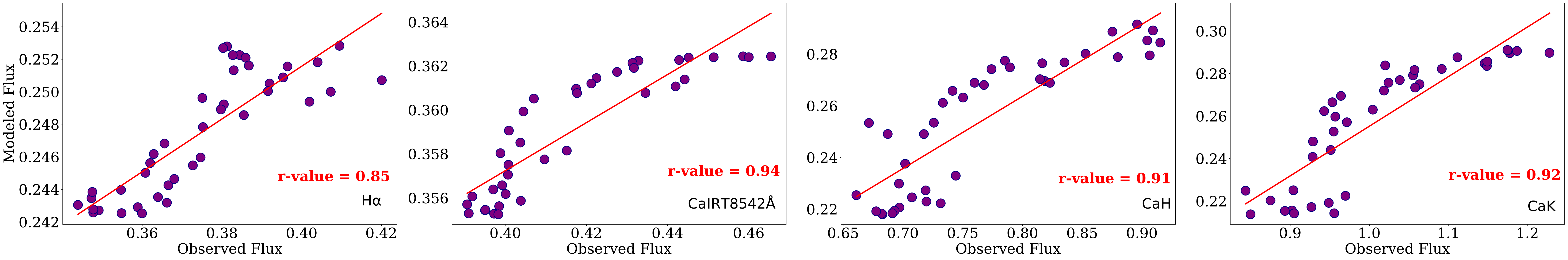}
    \caption{Flux correlations between the observed and the modeled values. The red line show the linear fit for the given chromospheric emissions. The r-values indicate spearman correlation coefficients for the given measurements. }
    \label{flux_corr}
\end{figure*}

\subsection{Chromospheric emissions}\label{sec:Chromo}
We measured the line-core emission fluxes in the 1-\AA\,region centered around the line cores of \cahk, \Halpha, and \cairtt. Relative flux measurements are obtained by integrating the area between the spectrum and the zero point. Unit conversions and further discussion for radiative loss calculations can be found in \citet{Adebali2025} and \citet{Jarvinen2025}. After calculating the emission fluxes (Fig.\ref{Chromo_figs}), we fitted a sinusoidal function to the phase-dependent relative flux values via the $\chi^{2}$ routine of \texttt{scipy}\ package \citep{SciPy2020-NMeth}. Based on our fits, we determined that the maximum emission excess for \cahk\,which is about $\simeq$ 30\%. These two emission indicators are the ones with the highest modulation amplitudes. \Halpha\ fluxes follow them with a much lower modulation rate with about 5\%. The least modulation rate is determined for \cairtt, where the amplitude changes over a rotation with a rate about 2\% of the minimum value of the calculated fluxes. At the same time, we note that in all four cases the simulated activity signal is weaker than that of the \lama, with the biggest difference coming from the \cahk, then \Halpha, and finally \cairtt. This differential response in chromospheric lines is consistent with that reported by \citet{Pietrow2024_harps} and the references therein. The discrepancy between the shape of each curve and the original data, most clearly visible in Fig.~\ref{Chromo_figs}, correlates with the degree to which the average penumbral profile resembles that of the umbra (See Fig.~\ref{fig:avgprof}). The largest scatter is found in the H$\alpha$ line, where the penumbral profile is nearly indistinguishable from that of the quiet Sun having the lowest corelation coefficient (see Fig. \ref{flux_corr}). This significantly reduces the contrast of the penumbral contribution on the stellar surface, effectively giving greater weight to the darker umbral component and thereby altering the recovered line profile. The \cairtt\ umbral profile has a similar issue, where it has an equal depth to the surrounding quiet Sun, but also shows an asymmetrical broadening. This broadening could introduce an asymmetry in response to penumbral patches on the left and right sides of the disk, which indeed seems to be the case in the second panel of Fig.~\ref{Chromo_figs}, where a reduced response is seen when the spot is on the right side of the disk.

The correlation between the integrated flux of the model and the observed spectra is shown in Fig.~\ref{flux_corr}. A clear correlation is present at low activity levels, while a distinct deviation appears at the highest values. This behavior likely reflects our assumption where large spots in the Doppler map are treated as monolithic structures. Interestingly, the \Halpha\,line does not seem to show this break, potentially making it the best line for this type of modeling.

\subsection{Radial velocities}\label{sec:RV}
\begin{figure*}
\includegraphics[width=\textwidth]{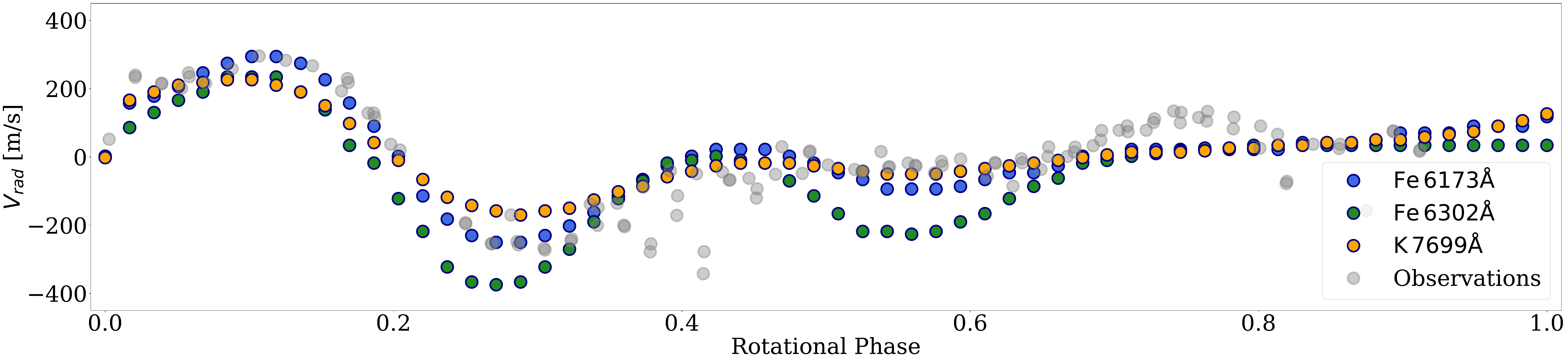}
\caption{RV evolution over the rotational phase. The blue, green and orange circles indicate $\rm Fe\,${\sc i }$ 6173\,$\AA, $\rm Fe\,${\sc i }$ 6302\,$\AA and $\rm K\,${\sc i }$7699\,$\AA\ respectively. The gray dots show the actual observations.}
\label{RVs_plot}
\end{figure*}

While computing the RV data points, we used the null-spectrum calculated by NESSI as a model spectrum and then obtained the RV points by using the cross-correlation function (CCF). Since the stellar parameters are the same for each spectrum, the only difference for each individual line should come from the surface structures imprinting on line core shifts and thus on RV values. As shown in Fig. \ref{RVs_plot}, phase-dependent RV modulations create similar patterns with different amplitudes for the used photospheric lines. The highest modulation is calculated for the iron line at 6302~\AA\,with an amplitude of $\simeq$ 320 \ms, and the lowest modulation is obtained by K {\sc i} 7699~\AA\,at around $\simeq$200 \ms. We attribute these differences due to two different effects; chromaticity and sensitivity of the lines to the temperature changes, which are formed in the given wavelength regions. The first effect is widely discussed by \citet{Laure2025}, who explain that in the blue spectral regions, the activity effects on RVs are observed more prominently. Furthermore, a second modulation appears between phases 0.4 - 0.6 with an amplitude of 60\ms\,and 120\ms\,for Fe 6173\AA\,and Fe 6302\AA\,respectively. The K 7699\AA\,shows a smaller amplitude of $\simeq$20\ms. 

%---------------------------------- Fig. 4

\section{Discussion} \label{sec:Discussion}
Using the NESSI code, we constructed a toy model of \lama\,based on the Doppler imaging map of \citet{Adebali2025}. We synthesized disk-integrated spectra from this surface map and compared them to the actual observations of the star that are simultaneously obtained with the used Doppler images. This allowed us to evaluate the origins of the chromospheric emissions of \lama\,and provide a solar-stellar connection in a different perspective.

Upon examining the activity modulation (as shown in Fig.~\ref{Chromo_figs}), we find that the overall variability patterns are qualitatively well reproduced with spot related activity alone despite the mismatches in the absolute flux levels and its modulation amplitude over one rotation. As these two stars demonstrate very different properties and we are only matching their temperature differences on their photospheres, this is very much in line with the expectations. On the other hand, \lama\,is very active in its chromospheres which already categorizes this system in the group of RS\,CVn type binaries. The fact that \cahk\,lines shows the highest modulation both in actual observations and in our toy model indicates that upper chromospheres of both stars are heated in relatively the same way, although the column density and the stratification of their are atmospheres are vastly different. As shown in Fig.~\ref{Chromo_figs}, the actual observations demonstrate a modulation amplitude of 35\% in average for the upper chromosphere and it is slightly higher than our results (30\%). The percentile difference for \Halpha\,is about 25\% in the observations, and for \cairtt\,this value is about 20\%. As it is shown at the upper chromosphere, the flux amplitudes inferred from these lines point out that \lama\,has more efficient heating in the lower and middle chromosphere than our toy model as well. In addition to that, the assumption of monolithic spots may be incorrect, and these regions may in fact be composed of smaller spots and surrounding plages, including possible complex small-scale interaction of magnetic fields, which would lower the contrast but raise the activity signal. 

The RV modulations are more closely aligned with each other when compared with the chromospheric variations, even though the spectra originate from different spots. Concurrently, this is expected since the RV variations depend more strongly on the filling factors of the spots as well as the spot temperatures. The temperature contrast, together with the size and the locations of the spots cause asymmetries for the observed parts of stellar-disk, this effect creates considerable difficulty while obtaining RV values \citep[see e.g.,][]{Boisse2011}. Although we compare different spots with different temperature contrasts for both stars, since we use the same surface structure from the Doppler imaging map, the observations are correlated better with the results from our toy model. However, when we examine the individual lines, the details become more prominent. The amplitude of the potassium line is about half that of the iron lines, which we interpret as the chromaticity effect on the RVs \citep[][]{Laure2025}. The average maximum RV amplitude calculated from these three lines is $\simeq$250\ms\,which is lower than the observed value of $\simeq$300 \ms\,\citep{Adebali2025}. We note that the RV values observed by \citet{Adebali2025} were obtained on the basis of 500 lines in a region between 4800 and 5400 \AA. Therefore, although the surface structures were introduced in the same way as in the Doppler images, the RV modulation based on our model is about 20\% less than the observed values, which is an underestimation in contrast to the model side. However, the RV observations are also affected by instrumental noise, and different surface effects (e.g., convective blue-shift) that are not possible to resolve in Doppler image of \lama.

In summary, from the chromospheric indices, we find a good match in shape but an underestimation in magnitude. From the photospheric RVs we find the same match in pattern, but underestimation in magnitude. This implies that our spots do not adequately reproduce the active regions on \lama. Also, perhaps contradictory, it expects darker spots in the photosphere and more active areas in the chromosphere. However, based on solar observations, we know that plages can be close to invisible at disk center when seen in the photosphere, while presenting as bright in the chromosphere. 

\section{Conclusion}
This experiment made use of publicly available tools to construct synthetic spectra representative of a G8-type giant star. However, present stellar-atmosphere modeling capabilities do not yet permit the computation of spectra from a realistic atmospheric model tailored to this star. As a result, solar spectra were used as a proxy. Although many of the observed spectral properties were reproduced, several discrepancies remain between the synthetic and observed spectra, which likely arise from differences in the atmospheric structure and parameters between the Sun and \lama, and the absence of plages and other bright structures in our model. However, the modulation of the RVs and activity indices is well aligned with the actual and simultaneous observations. This leads to the conclusion that, despite their limitations, toy models such as the one presented here retain significant diagnostic value. In particular, our results indicate that only spot spectra can account for the significant portion of the relative modulation of chromospheric emissions at different layers for this system, without requiring the inclusion of bright plages. However, at the same time, such features might be needed to better reproduce the magnitude of these modulations. Especially, for the actual observations of \lama\,the spots seem much more efficient for creating the mechanism of  heating the chromospheric layers than the sunspot observations we used for this experiment. Also, for the first time, we quantify that the main chromospheric emission source of \lama\,is its huge spot regions, which are followed through its Doppler images.

Future work will focus on expanding the analysis to a broader sample of stars, including solar-like and fast-rotating systems \citep[e.g., EK Dra;][]{Jarvinen2018, Gorgei26} and more RS\,CVn type systems \citep[e.g., HR 7275;][]{Adebali2026}, with developing and more advanced spectral-synthesis methods using more lines and a more realistic atmosphere.

%% Please use the acknowledgment and contribution environments. This will 
%% be anonomyized when the "anonymous" style option is used. 
\begin{acknowledgements}
We thank an anonymous referee for their constructive comments that significantly improved the quality of this work.
AP is supported by the Deut\-sche For\-schungs\-ge\-mein\-schaft (DFG) project number PI 2102/1-1.
\"OA acknowledges the insightful critiques from D. Gruner, J. Alvarado-Gomez, M. Weber and K. G. Strassmeier.
In this work, we heavily used \texttt{python3} libraries; \texttt{astropy}\, \citep{astropy:2013, astropy:2018, astropy:2022},\, \texttt{numpy}\, \citep{numpyharris2020}\, \texttt{scipy}\, \citep{SciPy2020-NMeth}, and ISPy \citep{DiazBaso2021}.
DeepL Write was used in copy editing (spelling, grammar, and readability) of the manuscript.
\end{acknowledgements}

\bibliographystyle{aa}
\bibliography{aa59430-26.bib}

@ARTICLE{Han2025,
       author = {{Han}, Henggeng and {Wang}, Song and {Li}, Xue and {Zheng}, Chuanjie and {Liu}, Jifeng},
        title = "{Impact of Spectral Resolution on S-index and Its Application to Spectroscopic Surveys}",
      journal = {\apj},
         year = 2025,
        month = may,
       volume = {984},
       number = {1},
          eid = {2},
        pages = {2},
          doi = {10.3847/1538-4357/adc600},
archivePrefix = {arXiv},
       eprint = {2503.20165},
 primaryClass = {astro-ph.SR},
       adsurl = {https://ui.adsabs.harvard.edu/abs/2025ApJ...984....2H}
}

@ARTICLE{Pietrow2026,
       author = {{Pietrow}, Alexander G.~M.},
        title = "{HelioSpectrotron 5000: an interactive solar atlas}",
      journal = {The Open Journal of Astrophysics},
         year = 2026,
        month = feb,
       volume = {9},
        pages = {58273},
          doi = {10.33232/001c.158273},
archivePrefix = {arXiv},
       eprint = {2602.20101},
 primaryClass = {astro-ph.SR},
       adsurl = {https://ui.adsabs.harvard.edu/abs/2026OJAp....958273P}
}

@ARTICLE{Neckel1984,
       author = {{Neckel}, H. and {Labs}, D.},
        title = "{The solar radiation between 3300 and 12500 {\r{A}}}",
      journal = {\solphys},
         year = 1984,
        month = feb,
       volume = {90},
       number = {2},
        pages = {205-258},
          doi = {10.1007/BF00173953},
       adsurl = {https://ui.adsabs.harvard.edu/abs/1984SoPh...90..205N}
}

@ARTICLE{Hanassi2025,
       author = {{Hanassi-Savari}, F. and {Pietrow}, A.~G.~M. and {Druett}, M.~K. and {Cretignier}, M. and {Ellwarth}, M.},
        title = "{Solar flux atlases: The new HARPS-N quiet Sun benchmark and continuum normalisation of the Ca II H \& K lines}",
      journal = {\aap},
         year = 2025,
        month = oct,
       volume = {702},
          eid = {A97},
        pages = {A97},
          doi = {10.1051/0004-6361/202451874},
archivePrefix = {arXiv},
       eprint = {2508.07912},
 primaryClass = {astro-ph.SR},
       adsurl = {https://ui.adsabs.harvard.edu/abs/2025A&A...702A..97H}
}

@INPROCEEDINGS{DiazBaso2021,
       author = {{D{\'\i}az Baso}, C.~J. and {Vissers}, G. and {Calvo}, F. and {Pietrow}, A.~G.~M. and {Yadav}, R. and {de la Cruz Rodr{\'\i}guez}, J. and {Zivadinovic}, L.},
        title = "{ISPy}",
    booktitle = {Zenodo Software package},
         year = 2021,
       volume = {56},
        month = oct,
    publisher = {Zenodo},
          eid = {5608441},
        pages = {5608441},
          doi = {10.5281/zenodo.5608441},
       adsurl = {https://ui.adsabs.harvard.edu/abs/2021zndo...5608441D}
}

@ARTICLE{Adebali2026,
       author = {{Adebali}, {\"O}. and {Weber}, M. and {Strassmeier}, K.~G. and {Ilyin}, I.~V. and {Steffen}, M. and {K{\H{o}}v{\'a}ri}, Zs.},
        title = "{Surface image and activity-corrected orbit of the RSCVn binary HR7275: Disentangling activity tracers}",
      journal = {\aap},
         year = 2026,
        month = feb,
       volume = {706},
          eid = {A179},
        pages = {A179},
          doi = {10.1051/0004-6361/202557717},
       adsurl = {https://ui.adsabs.harvard.edu/abs/2026A&A...706A.179A}
}

@ARTICLE{Adebali2025,
       author = {{Adebali}, {\"O}. and {Strassmeier}, K.~G. and {Ilyin}, I.~V. and {Weber}, M. and {Gruner}, D. and {K{\H{o}}v{\'a}ri}, Zs.},
        title = "{First Doppler image and starspot-corrected orbit for {\ensuremath{\lambda}} Andromedae: A multifaceted activity analysis}",
      journal = {\aap},
         year = 2025,
        month = mar,
       volume = {695},
          eid = {A89},
        pages = {A89},
          doi = {10.1051/0004-6361/202453073},
       adsurl = {https://ui.adsabs.harvard.edu/abs/2025A&A...695A..89A}
}

@article{astropy:2013,
Adsurl = {http://adsabs.harvard.edu/abs/2013A%26A...558A..33A},
Archiveprefix = {arXiv},
Author = {{Astropy Collaboration} and {Robitaille}, T.~P. and others},
Doi = {10.1051/0004-6361/201322068},
Eid = {A33},
Eprint = {1307.6212},
Journal = {\aap},
Month = oct,
Pages = {A33},
Primaryclass = {astro-ph.IM},
Title = {{Astropy: A community Python package for astronomy}},
Volume = 558,
Year = 2013
}

@ARTICLE{astropy:2018,
       author = {{Astropy Collaboration} and {Price-Whelan}, A.~M. and
         others},
        title = "{The Astropy Project: Building an Open-science Project and Status of the v2.0 Core Package}",
      journal = {\aj},
         year = 2018,
        month = sep,
       volume = {156},
       number = {3},
          eid = {123},
        pages = {123},
          doi = {10.3847/1538-3881/aabc4f},
archivePrefix = {arXiv},
       eprint = {1801.02634},
 primaryClass = {astro-ph.IM},
       adsurl = {https://ui.adsabs.harvard.edu/abs/2018AJ....156..123A}
}

@BOOK{Gray2022,
       author = {{Gray}, David F.},
        title = "{The observation and analysis of stellar photospheres}",
         year = 2022,
          doi = {10.1017/9781009082136},
       adsurl = {https://ui.adsabs.harvard.edu/abs/2022oasp.book.....G}
}

@ARTICLE{Pietrow2024_harps,
       author = {{Pietrow}, A.~G.~M. and {Cretignier}, M. and {Druett}, M.~K. and {Alvarado-G{\'o}mez}, J.~D. and {Hofmeister}, S.~J. and {Verma}, M. and {Kamlah}, R. and {Baratella}, M. and {Amazo-G{\'o}mez}, E.~M. and {Kontogiannis}, I. and {Dineva}, E. and {Warmuth}, A. and {Denker}, C. and {Poppenhaeger}, K. and {Andriienko}, O. and {Dumusque}, X. and {L{\"o}fdahl}, M.~G.},
        title = "{A comparative study of two X2.2 and X9.3 solar flares observed with HARPS-N. Reconciling Sun-as-a-star spectroscopy and high-spatial resolution solar observations in the context of the solar-stellar connection}",
      journal = {\aap},
         year = 2024,
        month = feb,
       volume = {682},
          eid = {A46},
        pages = {A46},
          doi = {10.1051/0004-6361/202347895},
archivePrefix = {arXiv},
       eprint = {2309.03373},
 primaryClass = {astro-ph.SR},
       adsurl = {https://ui.adsabs.harvard.edu/abs/2024A&A...682A..46P}
}

@ARTICLE{astropy:2022,
       author = {{Astropy Collaboration} and {Price-Whelan}, Adrian M. and others},
        title = "{The Astropy Project: Sustaining and Growing a Community-oriented Open-source Project and the Latest Major Release (v5.0) of the Core Package}",
      journal = {\apj},
         year = 2022,
        month = aug,
       volume = {935},
       number = {2},
          eid = {167},
        pages = {167},
          doi = {10.3847/1538-4357/ac7c74},
archivePrefix = {arXiv},
       eprint = {2206.14220},
 primaryClass = {astro-ph.IM},
       adsurl = {https://ui.adsabs.harvard.edu/abs/2022ApJ...935..167A}
}

@ARTICLE{JAG2016,
       author = {{Alvarado-G{\'o}mez}, J.~D. and {Hussain}, G.~A.~J. and others},
        title = "{Simulating the environment around planet-hosting stars. II. Stellar winds and inner astrospheres}",
      journal = {\aap},
         year = 2016,
        month = oct,
       volume = {594},
          eid = {A95},
        pages = {A95},
          doi = {10.1051/0004-6361/201628988},
archivePrefix = {arXiv},
       eprint = {1607.08405},
 primaryClass = {astro-ph.SR},
       adsurl = {https://ui.adsabs.harvard.edu/abs/2016A&A...594A..95A}
}

@ARTICLE{Cristo2025,
       author = {{Cristo}, E. and {Faria}, J.~P. and {Santos}, N.~C. and {Dethier}, W. and {Akinsanmi}, B. and {Barka}, A. and {Demangeon}, O. and {Lucero}, J.~P. and {Silva}, A.~M.},
        title = "{SOAPv4: A new step toward modeling stellar signatures in exoplanet research}",
      journal = {\aap},
         year = 2025,
        month = oct,
       volume = {702},
          eid = {A84},
        pages = {A84},
          doi = {10.1051/0004-6361/202555184},
archivePrefix = {arXiv},
       eprint = {2510.08319},
 primaryClass = {astro-ph.EP},
       adsurl = {https://ui.adsabs.harvard.edu/abs/2025A&A...702A..84C}
}

@ARTICLE{Petit2024,
       author = {{Petit dit de la Roche}, D.~J.~M. and {Chakraborty}, H. and others},
        title = "{Detection of faculae in the transit and transmission spectrum of WASP-69b}",
      journal = {\aap},
         year = 2024,
        month = dec,
       volume = {692},
          eid = {A83},
        pages = {A83},
          doi = {10.1051/0004-6361/202451740},
archivePrefix = {arXiv},
       eprint = {2410.18663},
 primaryClass = {astro-ph.EP},
       adsurl = {https://ui.adsabs.harvard.edu/abs/2024A&A...692A..83P}
}

@ARTICLE{Solanki2003,
       author = {{Solanki}, Sami K.},
        title = "{Sunspots: An overview}",
      journal = {\aapr},
         year = 2003,
        month = jan,
       volume = {11},
       number = {2-3},
        pages = {153-286},
          doi = {10.1007/s00159-003-0018-4},
       adsurl = {https://ui.adsabs.harvard.edu/abs/2003A&ARv..11..153S}
}

@ARTICLE{Cretignier2024,
       author = {{Cretignier}, M. and {Pietrow}, A.~G.~M. and {Aigrain}, S.},
        title = "{Stellar surface information from the Ca II H\&K lines - I. Intensity profiles of the solar activity components}",
      journal = {\mnras},
         year = 2024,
        month = jan,
       volume = {527},
       number = {2},
        pages = {2940-2962},
          doi = {10.1093/mnras/stad3292},
archivePrefix = {arXiv},
       eprint = {2310.15926},
 primaryClass = {astro-ph.SR},
       adsurl = {https://ui.adsabs.harvard.edu/abs/2024MNRAS.527.2940C}
}

@ARTICLE{Rodberg1966,
       author = {{R{\"o}dberg}, Harald},
        title = "{Umbra Intensity of a Large Sunspot}",
      journal = {\nat},
         year = 1966,
        month = jul,
       volume = {211},
       number = {5047},
        pages = {394-395},
          doi = {10.1038/211394a0},
       adsurl = {https://ui.adsabs.harvard.edu/abs/1966Natur.211..394R}
}

@ARTICLE{Emily2025,
       author = {{L{\"o}{\ss}nitz}, Emily Joe and {Pietrow}, Alexander G.~M. and {Chakraborty}, Hritam and {Verma}, Meetu and {Kontogiannis}, Ioannis and {Balthasar}, Horst and {Denker}, Carsten and {Lendl}, Monika},
        title = "{Differential rotation of solar {\ensuremath{\alpha}} sunspots and implications for stellar light curves}",
      journal = {\aap},
         year = 2025,
        month = nov,
       volume = {703},
          eid = {A187},
        pages = {A187},
          doi = {10.1051/0004-6361/202555654},
archivePrefix = {arXiv},
       eprint = {2508.08196},
 primaryClass = {astro-ph.SR},
       adsurl = {https://ui.adsabs.harvard.edu/abs/2025A&A...703A.187L}
}

@ARTICLE{Ellwarth2023,
       author = {{Ellwarth}, M. and {Sch{\"a}fer}, S. and {Reiners}, A. and {Zechmeister}, M.},
        title = "{The IAG spectral atlas of the spatially resolved Sun: Centre-to-limb observations}",
      journal = {\aap},
         year = 2023,
        month = may,
       volume = {673},
          eid = {A19},
        pages = {A19},
          doi = {10.1051/0004-6361/202245612},
archivePrefix = {arXiv},
       eprint = {2303.08205},
 primaryClass = {astro-ph.SR},
       adsurl = {https://ui.adsabs.harvard.edu/abs/2023A&A...673A..19E}
}

@ARTICLE{Scherrer2012,
       author = {{Scherrer}, P.~H. and {Schou}, J. and {Bush}, R.~I. and {Kosovichev}, A.~G. and {Bogart}, R.~S. and {Hoeksema}, J.~T. and {Liu}, Y. and {Duvall}, T.~L. and {Zhao}, J. and {Title}, A.~M. and {Schrijver}, C.~J. and {Tarbell}, T.~D. and {Tomczyk}, S.},
        title = "{The Helioseismic and Magnetic Imager (HMI) Investigation for the Solar Dynamics Observatory (SDO)}",
      journal = {\solphys},
         year = 2012,
        month = jan,
       volume = {275},
       number = {1-2},
        pages = {207-227},
          doi = {10.1007/s11207-011-9834-2},
       adsurl = {https://ui.adsabs.harvard.edu/abs/2012SoPh..275..207S}
}

@ARTICLE{Fontenla1993,
       author = {{Fontenla}, J.~M. and {Avrett}, E.~H. and {Loeser}, R.},
        title = "{Energy Balance in the Solar Transition Region. III. Helium Emission in Hydrostatic, Constant-Abundance Models with Diffusion}",
      journal = {\apj},
         year = 1993,
        month = mar,
       volume = {406},
        pages = {319},
          doi = {10.1086/172443},
       adsurl = {https://ui.adsabs.harvard.edu/abs/1993ApJ...406..319F}
}

@ARTICLE{Pesnell,
       author = {{Pesnell}, W. Dean and {Thompson}, B.~J. and {Chamberlin}, P.~C.},
        title = "{The Solar Dynamics Observatory (SDO)}",
      journal = {\solphys},
         year = 2012,
        month = jan,
       volume = {275},
       number = {1-2},
        pages = {3-15},
          doi = {10.1007/s11207-011-9841-3},
       adsurl = {https://ui.adsabs.harvard.edu/abs/2012SoPh..275....3P}
}

@INPROCEEDINGS{Pietrow2024,
       author = {{Pietrow}, Alexander G.~M. and {Pastor Yabar}, Adur},
        title = "{Center-to-limb variation of spectral lines and their effect on full-disk observations}",
    booktitle = {Dynamics of Solar and Stellar Convection Zones and Atmospheres},
         year = 2024,
       editor = {{Getling}, Alexander V. and {Kitchatinov}, Leonid L.},
       series = {IAU Symposium},
       volume = {365},
        month = dec,
        pages = {389-393},
          doi = {10.1017/S174392132300501X},
archivePrefix = {arXiv},
       eprint = {2311.06200},
 primaryClass = {astro-ph.SR},
       adsurl = {https://ui.adsabs.harvard.edu/abs/2024IAUS..365..389P}
}

@ARTICLE{dewilde2025,
       author = {{De Wilde}, M. and {Pietrow}, A.~G.~M. and {Druett}, M.~K. and {Pastor Yabar}, A. and {Koza}, J. and {Kontogiannis}, I. and {Andriienko}, O. and {Berlicki}, A. and {Brunvoll}, A.~R. and {de la Cruz Rodr{\'\i}guez}, J. and {Thoen Faber}, J. and {Joshi}, R. and {Kuridze}, D. and {N{\'o}brega-Siverio}, D. and {Rouppe van der Voort}, L.~H.~M. and {Ryb{\'a}k}, J. and {Scullion}, E. and {Silva}, A.~M. and {Vashalomidze}, Z. and {Vicente Ar{\'e}valo}, A. and {Wi{\'s}niewska}, A. and {Yadav}, R. and {Zaqarashvili}, T.~V. and {Zbinden}, J. and {{\O}yre}, E.~S.},
        title = "{Synthesizing Sun-as-a-star flare spectra from high-resolution solar observations}",
      journal = {\aap},
         year = 2025,
        month = aug,
       volume = {700},
          eid = {A275},
        pages = {A275},
          doi = {10.1051/0004-6361/202554870},
archivePrefix = {arXiv},
       eprint = {2507.07967},
 primaryClass = {astro-ph.SR},
       adsurl = {https://ui.adsabs.harvard.edu/abs/2025A&A...700A.275D}
}

@ARTICLE{Gorgei26,
       author = {{G{\"o}rgei}, A. and {Kriskovics}, L. and {Vida}, K. and {Seli}, B. and {Ol{\'a}h}, K. and {S{\'a}gi}, P. and {B{\'o}di}, A. and {J{\"a}rvinen}, S.~P. and {Strassmeier}, K.~G. and {P{\'a}l}, A. and {K{\H{o}}v{\'a}ri}, Zs.},
        title = "{Magnetic activity on the young Sun: A case study of EK Draconis}",
      journal = {\aap},
         year = 2026,
        month = jan,
       volume = {706},
          eid = {A54},
        pages = {A54},
          doi = {10.1051/0004-6361/202556949},
archivePrefix = {arXiv},
       eprint = {2512.03830},
 primaryClass = {astro-ph.SR},
       adsurl = {https://ui.adsabs.harvard.edu/abs/2026A&A...706A..54G}
}

@ARTICLE{Jarvinen2018,
       author = {{J{\"a}rvinen}, S.~P. and {Strassmeier}, K.~G. and {Carroll}, T.~A. and others},
        title = "{Mapping EK Draconis with PEPSI. Possible evidence for starspot penumbrae}",
      journal = {\aap},
         year = 2018,
        month = dec,
       volume = {620},
          eid = {A162},
        pages = {A162},
          doi = {10.1051/0004-6361/201833496},
archivePrefix = {arXiv},
       eprint = {1812.03675},
 primaryClass = {astro-ph.SR},
       adsurl = {https://ui.adsabs.harvard.edu/abs/2018A&A...620A.162J}
}

@ARTICLE{Kuckein2021,
       author = {{Kuckein}, C. and {Balthasar}, H. and {Quintero Noda}, C. and {Diercke}, A. and {Trelles Arjona}, J.~C. and {Ruiz Cobo}, B. and {Felipe}, T. and {Denker}, C. and {Verma}, M. and {Kontogiannis}, I. and {Sobotka}, M.},
        title = "{Multiple Stokes I inversions for inferring magnetic fields in the spectral range around Cr I 5782 {\r{A}}}",
      journal = {\aap},
         year = 2021,
        month = sep,
       volume = {653},
          eid = {A165},
        pages = {A165},
          doi = {10.1051/0004-6361/202140596},
archivePrefix = {arXiv},
       eprint = {2107.11116},
 primaryClass = {astro-ph.SR},
       adsurl = {https://ui.adsabs.harvard.edu/abs/2021A&A...653A.165K}
}

@ARTICLE{Bortle2021,
       author = {{Bortle}, Anna and {Fausey}, Hallie and {Ji}, Jinbiao and {Dodson-Robinson}, Sarah and {Ramirez Delgado}, Victor and {Gizis}, John},
        title = "{A Gaussian Process Regression Reveals No Evidence for Planets Orbiting Kapteyn's Star}",
      journal = {\aj},
         year = 2021,
        month = may,
       volume = {161},
       number = {5},
          eid = {230},
        pages = {230},
          doi = {10.3847/1538-3881/abec89},
archivePrefix = {arXiv},
       eprint = {2103.02709},
 primaryClass = {astro-ph.EP},
       adsurl = {https://ui.adsabs.harvard.edu/abs/2021AJ....161..230B}
}

@ARTICLE{Martin2017,
       author = {{Martin}, J. and {Fuhrmeister}, B. and {Mittag}, M. and {Schmidt}, T.~O.~B. and {Hempelmann}, A. and {Gonz{\'a}lez-P{\'e}rez}, J.~N. and {Schmitt}, J.~H.~M.~M.},
        title = "{The Ca II infrared triplet's performance as an activity indicator compared to Ca II H and K. Empirical relations to convert Ca II infrared triplet measurements to common activity indices}",
      journal = {\aap},
         year = 2017,
        month = sep,
       volume = {605},
          eid = {A113},
        pages = {A113},
          doi = {10.1051/0004-6361/201630298},
       adsurl = {https://ui.adsabs.harvard.edu/abs/2017A&A...605A.113M}
}

@ARTICLE{Kovari2016,
       author = {{K{\H{o}}v{\'a}ri}, Zs. and {K{\"u}nstler}, A. and {Strassmeier}, K.~G. and {Carroll}, T.~A. and {Weber}, M. and {Kriskovics}, L. and {Ol{\'a}h}, K. and {Vida}, K. and {Granzer}, T.},
        title = "{Time-series Doppler images and surface differential rotation of the effectively single, rapidly rotating K-giant KU Pegasi}",
      journal = {\aap},
         year = 2016,
        month = nov,
       volume = {596},
          eid = {A53},
        pages = {A53},
          doi = {10.1051/0004-6361/201628425},
archivePrefix = {arXiv},
       eprint = {1609.00196},
 primaryClass = {astro-ph.SR},
       adsurl = {https://ui.adsabs.harvard.edu/abs/2016A&A...596A..53K}
}

@ARTICLE{Mittag2013,
       author = {{Mittag}, M. and {Schmitt}, J.~H.~M.~M. and {Schr{\"o}der}, K.-P.},
        title = "{Ca II H+K fluxes from S-indices of large samples: a reliable and consistent conversion based on PHOENIX model atmospheres}",
      journal = {\aap},
         year = 2013,
        month = jan,
       volume = {549},
          eid = {A117},
        pages = {A117},
          doi = {10.1051/0004-6361/201219868},
       adsurl = {https://ui.adsabs.harvard.edu/abs/2013A&A...549A.117M}
}

@ARTICLE{Moulds2013,
       author = {{Moulds}, V.~E. and {Watson}, C.~A. and {Bonfils}, X. and others},
        title = "{Finding exoplanets orbiting young active stars - I. Technique}",
      journal = {\mnras},
         year = 2013,
        month = apr,
       volume = {430},
       number = {3},
        pages = {1709-1721},
          doi = {10.1093/mnras/sts709},
archivePrefix = {arXiv},
       eprint = {1212.5922},
 primaryClass = {astro-ph.EP},
       adsurl = {https://ui.adsabs.harvard.edu/abs/2013MNRAS.430.1709M}
}

@ARTICLE{Simpson2022,
       author = {{Simpson}, Emilie R. and {Fetherolf}, Tara and {Kane}, Stephen R. and {Li}, Zhexing and {Pepper}, Joshua and {Mo{\v{c}}nik}, Teo},
        title = "{Revisiting BD-06 1339b: A Likely False Positive Caused by Stellar Activity}",
      journal = {\aj},
         year = 2022,
        month = may,
       volume = {163},
       number = {5},
          eid = {215},
        pages = {215},
          doi = {10.3847/1538-3881/ac5d41},
archivePrefix = {arXiv},
       eprint = {2203.06191},
 primaryClass = {astro-ph.EP},
       adsurl = {https://ui.adsabs.harvard.edu/abs/2022AJ....163..215S}
}

@ARTICLE{Reiners2009,
       author = {{Reiners}, A.},
        title = "{Activity-induced radial velocity jitter in a flaring M dwarf}",
      journal = {\aap},
         year = 2009,
        month = may,
       volume = {498},
       number = {3},
        pages = {853-861},
          doi = {10.1051/0004-6361/200810257},
archivePrefix = {arXiv},
       eprint = {0903.2661},
 primaryClass = {astro-ph.SR},
       adsurl = {https://ui.adsabs.harvard.edu/abs/2009A&A...498..853R}
}

@ARTICLE{Solanki2004,
       author = {{Solanki}, S.~K. and {Unruh}, Y.~C.},
        title = "{Spot sizes on Sun-like stars}",
      journal = {\mnras},
         year = 2004,
        month = feb,
       volume = {348},
       number = {1},
        pages = {307-315},
          doi = {10.1111/j.1365-2966.2004.07368.x},
archivePrefix = {arXiv},
       eprint = {astro-ph/0311310},
 primaryClass = {astro-ph},
       adsurl = {https://ui.adsabs.harvard.edu/abs/2004MNRAS.348..307S}
}

@ARTICLE{Canocchi2024,
       author = {{Canocchi}, G. and {Lind}, K. and {Lagae}, C. and {Pietrow}, A.~G.~M. and {Amarsi}, A.~M. and {Kiselman}, D. and {Andriienko}, O. and {Hoeijmakers}, H.~J.},
        title = "{3D non-LTE modeling of the stellar center-to-limb variation for transmission spectroscopy studies. Na I D and K I resonance lines in the Sun}",
      journal = {\aap},
         year = 2024,
        month = mar,
       volume = {683},
          eid = {A242},
        pages = {A242},
          doi = {10.1051/0004-6361/202347858},
archivePrefix = {arXiv},
       eprint = {2312.05078},
 primaryClass = {astro-ph.SR},
       adsurl = {https://ui.adsabs.harvard.edu/abs/2024A&A...683A.242C}
}

@ARTICLE{Pietrow2023,
       author = {{Pietrow}, A.~G.~M. and {Kiselman}, D. and {Andriienko}, O. and {Petit dit de la Roche}, D.~J.~M. and {D{\'\i}az Baso}, C.~J. and {Calvo}, F.},
        title = "{Center-to-limb variation of spectral lines and continua observed with SST/CRISP and SST/CHROMIS}",
      journal = {\aap},
         year = 2023,
        month = mar,
       volume = {671},
          eid = {A130},
        pages = {A130},
          doi = {10.1051/0004-6361/202244811},
archivePrefix = {arXiv},
       eprint = {2212.03991},
 primaryClass = {astro-ph.SR},
       adsurl = {https://ui.adsabs.harvard.edu/abs/2023A&A...671A.130P}
}

@ARTICLE{Pietrow2025,
       author = {{Pietrow}, A.~G.~M. and {Kuckein}, C. and {Verma}, M. and {Denker}, C. and {Trelles Arjona}, J.~C. and {Kamlah}, R. and {Poppenh{\"a}ger}, K.},
        title = "{Center-to-limb variations of the He I 10 830 {\r{A}} triplet}",
      journal = {\aap},
         year = 2026,
        month = jan,
       volume = {705},
          eid = {A116},
        pages = {A116},
          doi = {10.1051/0004-6361/202557267},
archivePrefix = {arXiv},
       eprint = {2511.14331},
 primaryClass = {astro-ph.SR},
       adsurl = {https://ui.adsabs.harvard.edu/abs/2026A&A...705A.116P}
}

@ARTICLE{Scharmer08,
       author = {{Scharmer}, G.~B. and {Narayan}, G. and {Hillberg}, T. and
         {de la Cruz Rodriguez}, J. and {L{\"o}fdahl}, M.~G. and {Kiselman}, D. and
         {S{\"u}tterlin}, P. and {van Noort}, M. and {Lagg}, A.},
        title = "{CRISP Spectropolarimetric Imaging of Penumbral Fine Structure}",
      journal = {\apj},
         year = "2008",
        month = "Dec",
       volume = {689},
       number = {1},
        pages = {L69},
          doi = {10.1086/595744},
archivePrefix = {arXiv},
       eprint = {0806.1638},
 primaryClass = {astro-ph},
       adsurl = {https://ui.adsabs.harvard.edu/abs/2008ApJ...689L..69S}
}

@INPROCEEDINGS{Scharmer17,
       author = {{Scharmer}, Goran},
        title = "{SST/CHROMIS: a new window to the solar chromosphere}",
    booktitle = {SOLARNET IV: The Physics of the Sun from the Interior to the Outer Atmosphere},
         year = 2017,
        month = jan,
          eid = {85},
        pages = {85},
       adsurl = {https://ui.adsabs.harvard.edu/abs/2017psio.confE..85S}
}

@ARTICLE{Santos2025,
       author = {{Santos}, N.~C. and {Cabral}, A. and {Leite}, I. and {Smette}, A. and {Abreu}, M. and {Alves}, D. and {Martins}, J.~H.~C. and {Monteiro}, M. and {Silva}, A. and {Wehbe}, B. and {Arancibia}, J. and {{\'A}vila}, G. and {Brillant}, S. and {C{\'a}rdenas}, C. and {Clara}, R. and {Gafeira}, R. and {Gaytan}, D. and {Lovis}, C. and {Miranda}, N. and {Moreno}, P. and {Oliveira}, A. and {Otarola}, A. and {Pepe}, F. and {Rojas}, P. and {Schmutzer}, R. and {Sosnowska}, D. and {van der Heyden}, P. and {Al Moulla}, K. and {Adibekyan}, V. and {Barka}, A. and {Barros}, S.~C.~C. and {Branco}, P. and {Cristo}, E. and {Damasceno}, Y. and {Demangeon}, O. and {Dethier}, W. and {Faria}, J.~P. and {Gomes da Silva}, J. and {Gon{\c{c}}alves}, E. and {Lucero}, J.~P. and {Rodrigues}, J. and {San Nicolas Martinez}, C. and {Santos}, {\^A}. and {Sousa}, S. and {Viana}, P.~T.~P.},
        title = "{PoET: the Paranal solar ESPRESSO Telescope}",
      journal = {The Messenger},
         year = 2025,
        month = mar,
       volume = {194},
        pages = {21-25},
          doi = {10.18727/0722-6691/5381},
       adsurl = {https://ui.adsabs.harvard.edu/abs/2025Msngr.194...21S}
}

@BOOK{Wallace1999,
       author = {{Wallace}, L. and {Livingston}, W.~C. and {Bernath}, P.~F. and {Ram}, R.~S.},
        title = "{An atlas of the sunspot umbral spectrum in the red and infrared from 8900 to 15,050 cm(-1) (6642 to 11,230 [angstroms]), revised}",
         year = 1999,
       adsurl = {https://ui.adsabs.harvard.edu/abs/1999asus.book.....W}
}

@INPROCEEDINGS{Scharmer2003,
       author = {{Scharmer}, Goran B. and {Bjelksjo}, Klas and others},
        title = "{The 1-meter Swedish solar telescope}",
    booktitle = {Innovative Telescopes and Instrumentation for Solar Astrophysics},
         year = 2003,
       editor = {{Keil}, Stephen L. and {Avakyan}, Sergey V.},
       series = {Society of Photo-Optical Instrumentation Engineers (SPIE) Conference Series},
       volume = {4853},
        month = feb,
        pages = {341-350},
          doi = {10.1117/12.460377},
       adsurl = {https://ui.adsabs.harvard.edu/abs/2003SPIE.4853..341S}
}

@ARTICLE{Smitha2025,
       author = {{Smitha}, H.~N. and {Shapiro}, Alexander I. and {Witzke}, Veronika and {Kostogryz}, Nadiia M. and {Unruh}, Yvonne C. and {Bhatia}, Tanayveer S. and {Cameron}, Robert and {Seager}, Sara and {Solanki}, Sami K.},
        title = "{First Calculations of Starspot Spectra Based on 3D Radiative Magnetohydrodynamics Simulations}",
      journal = {\apjl},
         year = 2025,
        month = jan,
       volume = {978},
       number = {1},
          eid = {L13},
        pages = {L13},
          doi = {10.3847/2041-8213/ad9aaa},
archivePrefix = {arXiv},
       eprint = {2411.14056},
 primaryClass = {astro-ph.SR},
       adsurl = {https://ui.adsabs.harvard.edu/abs/2025ApJ...978L..13S}
}

@ARTICLE{Avrett2015,
       author = {{Avrett}, E. and {Tian}, H. and {Landi}, E. and {Curdt}, W. and {W{\"u}lser}, J.-P.},
        title = "{Modeling the Chromosphere of a Sunspot and the Quiet Sun}",
      journal = {\apj},
         year = 2015,
        month = oct,
       volume = {811},
       number = {2},
          eid = {87},
        pages = {87},
          doi = {10.1088/0004-637X/811/2/87},
       adsurl = {https://ui.adsabs.harvard.edu/abs/2015ApJ...811...87A}
}

@ARTICLE{Chakraborty2024,
       author = {{Chakraborty}, H. and {Lendl}, M. and {Akinsanmi}, B. and others},
        title = "{SAGE: A tool for constraining the impacts of stellar activity on transmission spectroscopy}",
      journal = {\aap},
         year = 2024,
        month = may,
       volume = {685},
          eid = {A173},
        pages = {A173},
          doi = {10.1051/0004-6361/202347727},
archivePrefix = {arXiv},
       eprint = {2311.16864},
 primaryClass = {astro-ph.EP},
       adsurl = {https://ui.adsabs.harvard.edu/abs/2024A&A...685A.173C}
}

@ARTICLE{Boisse2011,
       author = {{Boisse}, I. and {Bouchy}, F. and {H{\'e}brard}, G. and {Bonfils}, X. and {Santos}, N. and {Vauclair}, S.},
        title = "{Disentangling between stellar activity and planetary signals}",
      journal = {\aap},
         year = 2011,
        month = apr,
       volume = {528},
          eid = {A4},
        pages = {A4},
          doi = {10.1051/0004-6361/201014354},
       adsurl = {https://ui.adsabs.harvard.edu/abs/2011A&A...528A...4B}
}

@ARTICLE{Donati1995,
       author = {{Donati}, J. -F. and {Henry}, G.~W. and {Hall}, D.~S.},
        title = "{Activity, rotation and evolution of the RS CVn system {\ensuremath{\lambda}} Andromedae.}",
      journal = {\aap},
         year = 1995,
        month = jan,
       volume = {293},
        pages = {107-126},
       adsurl = {https://ui.adsabs.harvard.edu/abs/1995A&A...293..107D}
}

@ARTICLE{Hall1972,
       author = {{Hall}, Douglas S.},
        title = "{A T Tauri-Like Star in the Eclipsing Binary RS Canum Venaticorum}",
      journal = {\pasp},
         year = 1972,
        month = apr,
       volume = {84},
       number = {498},
        pages = {323},
          doi = {10.1086/129291},
       adsurl = {https://ui.adsabs.harvard.edu/abs/1972PASP...84..323H}
}

@ARTICLE{numpyharris2020,
 title         = {Array programming with {NumPy}},
 author        = {Charles R. Harris and K. Jarrod Millman and St{\'{e}}fan J.
                 van der Walt and Ralf Gommers and Pauli Virtanen and David
                 Cournapeau and Eric Wieser and Julian Taylor and Sebastian
                 Berg and Nathaniel J. Smith and Robert Kern and Matti Picus
                 and Stephan Hoyer and Marten H. van Kerkwijk and Matthew
                 Brett and Allan Haldane and Jaime Fern{\'{a}}ndez del
                 R{\'{i}}o and Mark Wiebe and Pearu Peterson and Pierre
                 G{\'{e}}rard-Marchant and Kevin Sheppard and Tyler Reddy and
                 Warren Weckesser and Hameer Abbasi and Christoph Gohlke and
                 Travis E. Oliphant},
 year          = {2020},
 month         = sep,
 journal       = {Nature},
 volume        = {585},
 number        = {7825},
 pages         = {357--362},
 doi           = {10.1038/s41586-020-2649-2},
 publisher     = {Springer Science and Business Media {LLC}},
 url           = {https://doi.org/10.1038/s41586-020-2649-2}
}

@ARTICLE{Jarvinen2025,
       author = {{J{\"a}rvinen}, S.~P. and {Strassmeier}, K.~G.},
        title = "{A search for Maunder-minimum candidate stars}",
      journal = {\aap},
         year = 2025,
        month = jun,
       volume = {698},
          eid = {A93},
        pages = {A93},
          doi = {10.1051/0004-6361/202554111},
archivePrefix = {arXiv},
       eprint = {2504.19670},
 primaryClass = {astro-ph.SR},
       adsurl = {https://ui.adsabs.harvard.edu/abs/2025A&A...698A..93J}
}

@ARTICLE{Laure2025,
       author = {{Larue}, P. and {Delfosse}, X. and {Carmona}, A. and {Meunier}, N. and {Artigau}, {\'E}. and {Bellotti}, S. and {Charpentier}, P. and {Moutou}, C. and {Donati}, J.-F. and {Boisse}, I. and {Forveille}, T. and {Arnold}, L. and {Bourrier}, V. and {Bonfils}, X. and {Cadieux}, C. and {Chomez}, A. and {Cook}, N. and {Cortes Zuleta}, P. and {Cristofari}, P. and {Diaz}, R. and {Doyon}, R. and {Grouffal}, S. and {Hara}, N. and {Heidari}, N. and {H{\'e}brard}, G. and {Kiefer}, F. and {Mignon}, L. and {Maurel}, A. and {Morin}, J. and {Petit}, A. and {Petit}, P. and {Santerne}, A. and {Santos}, N. and {Segransan}, D. and {Serrano Bell}, J. and {Vivien}, H.~G.},
        title = "{Chromaticity of stellar activity in radial velocities: Anti-correlated families of lines on the M dwarf EV Lac with SPIRou and SOPHIE}",
      journal = {\aap},
         year = 2025,
        month = sep,
       volume = {701},
          eid = {A216},
        pages = {A216},
          doi = {10.1051/0004-6361/202554963},
archivePrefix = {arXiv},
       eprint = {2509.17911},
 primaryClass = {astro-ph.SR},
       adsurl = {https://ui.adsabs.harvard.edu/abs/2025A&A...701A.216L}
}

@ARTICLE{Fionnagain2021,
       author = {{{\'O} Fionnag{\'a}in}, D. and {Vidotto}, A.~A. and {Petit}, P. and {Neiner}, C. and {Manchester}, W., IV and {Folsom}, C.~P. and {Hallinan}, G.},
        title = "{{\ensuremath{\lambda}} And: a post-main-sequence wind from a solar-mass star}",
      journal = {\mnras},
         year = 2021,
        month = jan,
       volume = {500},
       number = {3},
        pages = {3438-3453},
          doi = {10.1093/mnras/staa3468},
archivePrefix = {arXiv},
       eprint = {2011.02406},
 primaryClass = {astro-ph.SR},
       adsurl = {https://ui.adsabs.harvard.edu/abs/2021MNRAS.500.3438O}
}

@ARTICLE{Parks2021,
       author = {{Parks}, J.~R. and {White}, R.~J. and {Baron}, F. and {Monnier}, J.~D. and {Kloppenborg}, B. and {Henry}, G.~W. and {Schaefer}, G. and {Che}, X. and {Pedretti}, E. and {Thureau}, N. and {Zhao}, M. and {ten Brummelaar}, T. and {McAlister}, H. and {Ridgway}, S.~T. and {Turner}, N. and {Sturmann}, J. and {Sturmann}, L.},
        title = "{Interferometric Imaging of {\ensuremath{\lambda}} Andromedae: Evidence of Starspots and Rotation}",
      journal = {\apj},
         year = 2021,
        month = may,
       volume = {913},
       number = {1},
          eid = {54},
        pages = {54},
          doi = {10.3847/1538-4357/abb670},
       adsurl = {https://ui.adsabs.harvard.edu/abs/2021ApJ...913...54P}
}

@ARTICLE{SciPy2020-NMeth,
  author  = {Virtanen, Pauli and Gommers, Ralf and Oliphant, Travis E. and
            Haberland, Matt and Reddy, Tyler and Cournapeau, David and
            Burovski, Evgeni and Peterson, Pearu and Weckesser, Warren and
            Bright, Jonathan and {van der Walt}, St{\'e}fan J. and
            Brett, Matthew and Wilson, Joshua and Millman, K. Jarrod and
            Mayorov, Nikolay and Nelson, Andrew R. J. and Jones, Eric and
            Kern, Robert and Larson, Eric and Carey, C J and
            Polat, {\.I}lhan and Feng, Yu and Moore, Eric W. and
            {VanderPlas}, Jake and Laxalde, Denis and Perktold, Josef and
            Cimrman, Robert and Henriksen, Ian and Quintero, E. A. and
            Harris, Charles R. and Archibald, Anne M. and
            Ribeiro, Ant{\^o}nio H. and Pedregosa, Fabian and
            {van Mulbregt}, Paul and {SciPy 1.0 Contributors}},
  title   = {{{SciPy} 1.0: Fundamental Algorithms for Scientific
            Computing in Python}},
  journal = {Nature Methods},
  year    = {2020},
  volume  = {17},
  pages   = {261--272},
  adsurl  = {https://rdcu.be/b08Wh},
  doi     = {10.1038/s41592-019-0686-2},
}

@ARTICLE{Strassmeier2009,
       author = {{Strassmeier}, Klaus G.},
        title = "{Starspots}",
      journal = {\aapr},
         year = 2009,
        month = sep,
       volume = {17},
       number = {3},
        pages = {251-308},
          doi = {10.1007/s00159-009-0020-6},
       adsurl = {https://ui.adsabs.harvard.edu/abs/2009A&ARv..17..251S}
}

\clearpage
\onecolumn
\appendix
\section{Figures}

%---------------------------------- Fig. A2 (NOT a float)
\begin{center}
\centering
\includegraphics[width=\textwidth]{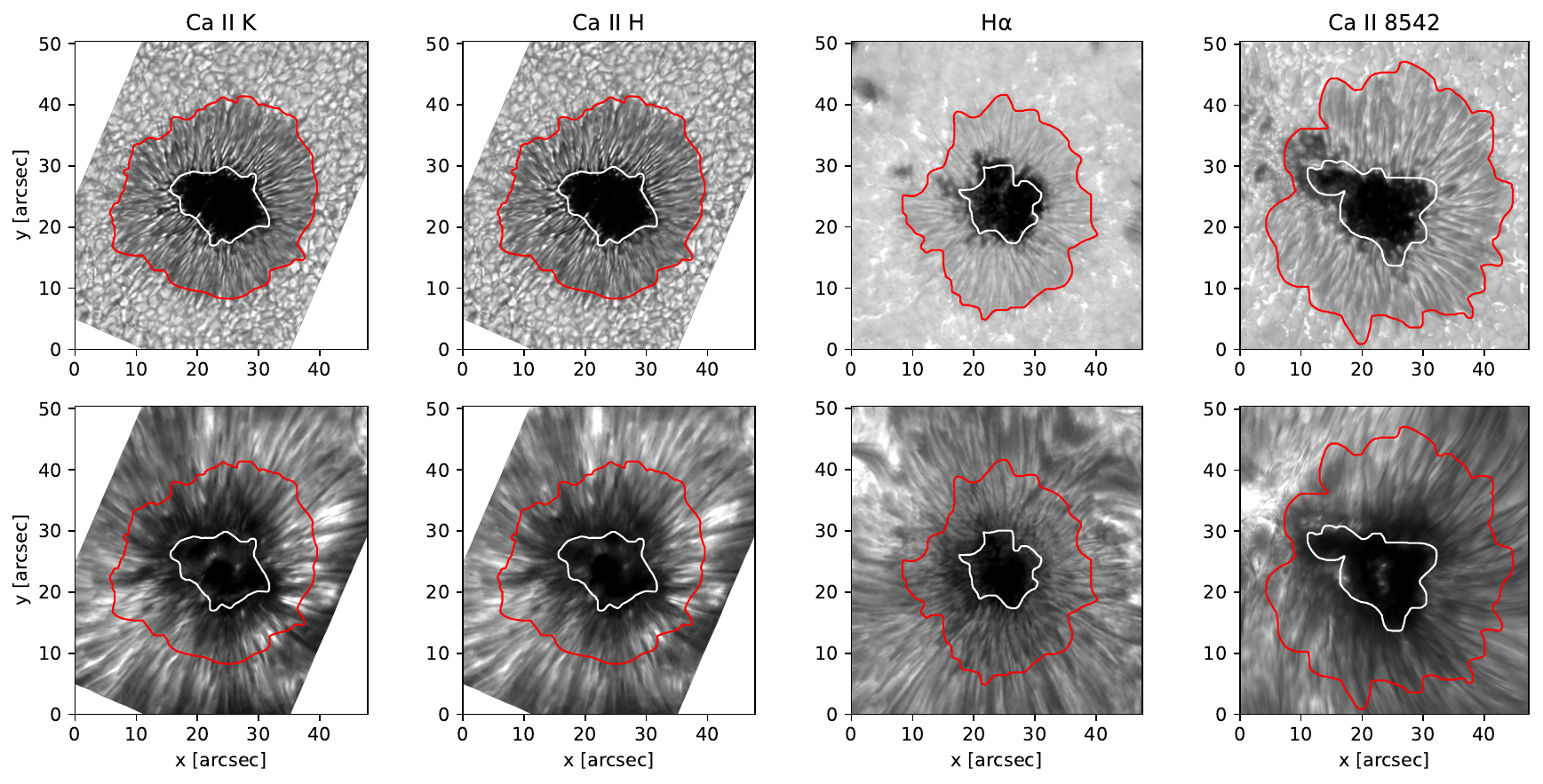}
\captionof{figure}{Sunspots used for the chromospheric-line analysis. \emph{Upper panel:} Line-wing images showing the (pseudo-)photospheric appearance of the spots. Contours derived from intensity thresholds outline the umbra (white) and penumbra (red). \emph{Lower panel:} The same fields of view shown in the respective line cores, probing the chromosphere at different heights. The same contours are overplotted to highlight the chromospheric contrast of the umbra, penumbra, and surrounding quiet Sun.}
\label{fig:spotsnessi}
\end{center}

%---------------------------------- Fig. A3 (NOT a float)
\begin{center}
\includegraphics[width=\textwidth]{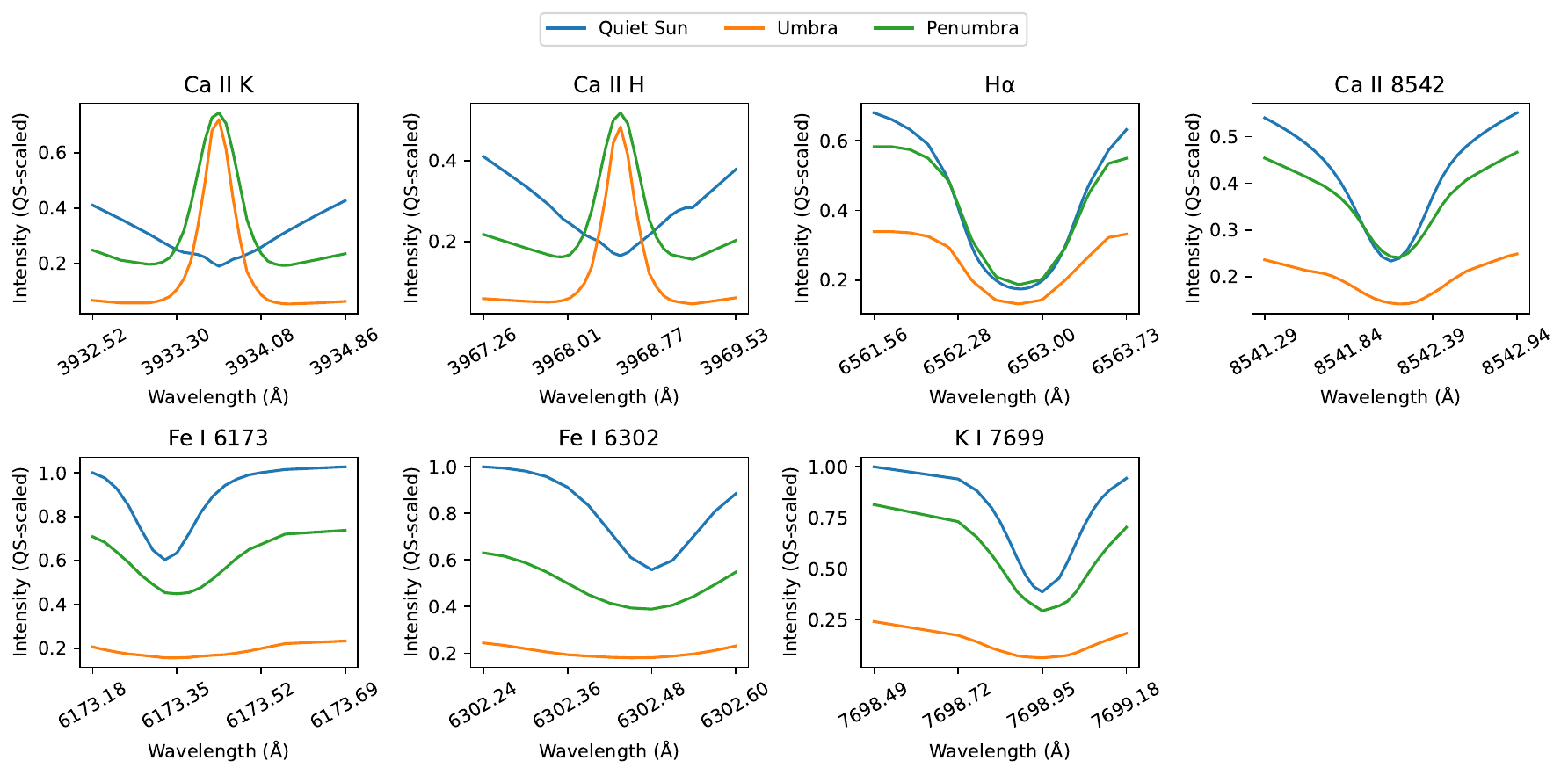}
\captionof{figure}{Resulting average profiles for quiet Sun (blue), penumbra (orange), and umbra (green) for eight spectral lines. For readability, all profiles are normalized to the their respective Quiet Sun profile.}
\label{fig:avgprof}
\end{center}

\newpage

\twocolumn
%------------------------------------- Fig. A4   
\begin{center}
\includegraphics[width=\textwidth/2]{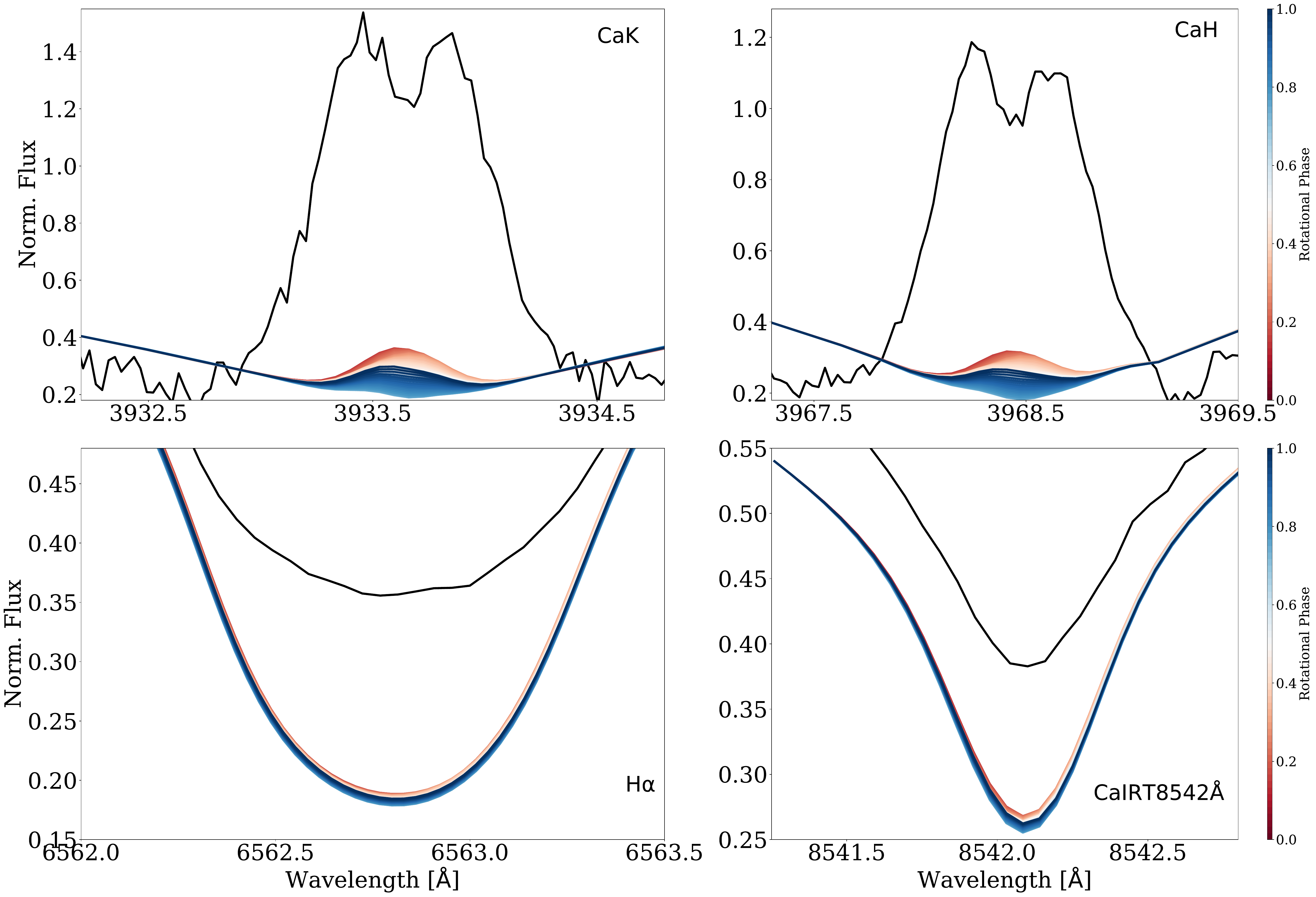}
\captionof{figure}{Color coded chromospheric emission lines compared with an observed profile at the median activity of \lama\,(black line).}
\label{Comp_lines}
\end{center}

\end{document}